\documentclass{WileyMSP-template}
\usepackage{amsmath,amssymb,amsbsy}
\usepackage{lineno}
\usepackage{hyperref}
\usepackage[numbers,square,sort&compress,super]{natbib}
\usepackage[T1]{fontenc} 
\usepackage{subcaption}
\modulolinenumbers[5]
\usepackage{tabularx}
\usepackage{wrapfig}
\usepackage{gensymb}
\usepackage{pifont}
\usepackage{tabu}
\usepackage{mathtools}
\usepackage{xcolor}
\usepackage{tikz}
\usepackage{array}
\usepackage{adjustbox}
\usepackage{wasysym}
\usepackage{blindtext}
\usepackage{geometry}
\usepackage{ragged2e}
\usepackage{amsmath}
\justifying
\newcommand{\baln}{B$_{x}$Al$_{1-x}$N }
\begin{document}

\pagestyle{fancy}
\rhead{\includegraphics[width=2.5cm]{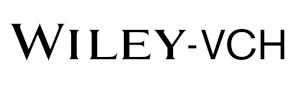}}

\title{ \emph{Ab-initio} Prediction of Ultra-Wide Band Gap B$_x$Al$_{1-x}$N Materials}

\maketitle

\author{Cody Milne*}
\author{Tathagata Biswas}
\author{Arunima K. Singh*}

\dedication{
}

\begin{affiliations}
Cody Milne, Tathagata Biswas, Arunima K. Singh\\
Arizona State University\\
975 S Myrtle Ave\\
Tempe, AZ 85281\\
Email: clmilne@asu.edu, arunimasingh@asu.edu, tbiswas3@asu.edu
\end{affiliations}


\begin{abstract}
  Ultra-wide bandgap (UWBG) materials are poised to play an important role in the future of power electronics. Devices made from UWBG materials are expected to operate at higher voltages, frequencies, and temperatures than current silicon and silicon carbide based devices; and can even lead to significant miniaturization of such devices. In the UWBG field, aluminum nitride and boron nitride have attracted great interest, however, the \baln alloys are much less studied. In this article, using first-principles simulations combining density-functional theory and the cluster expansion method we predict the crystal structure of \baln alloys. We find 17 ground state structures of \baln with formation energies between 0.11 and 0.25 eV/atom. All of these structures are found to be dynamically stable. The \baln structures are found to have predominantly a tetrahedral bonding environment, however, some structures exhibit $sp^2$ bonds similar to hexagonal BN. This work expands our knowledge of the structures, energies, and bonding in \baln which aids their synthesis, the innovation of lateral or vertical devices, and discovery of compatible dielectric and Ohmic contact materials. 
\end{abstract}

\twocolumn
\section{Introduction}
Ultra-wide bandgap (UWBG) semiconductors have recently emerged as an exciting class of materials due to their potential applications in power electronics, optoelectronics, and radio frequency devices. UWBG materials are defined as materials that have a bandgap larger than that of GaN (3.4 eV)\cite{uwbg-semiconductors}. The large bandgap of materials impacts many device performance parameters, such as thinner drift layers and lower specific on resistance, allowing significant miniaturization of devices like switches and transistors. Furthermore, UWBG materials are generally characterized by high bandgap, very large breakdown fields ($>10^{6}$ V/cm), high thermal conductivity, and reduced impact ionization rates and tunnelling due their high bandgaps and high mechanical strengths. AlN and BN are the largest gap group-III nitrides, and AlN has already been explored for application in the field of power electronics and in UV device applications\cite{Li2015, Nwigboji2015, Yu2021}. AlGaN alloys have also been used extensively in power electronics due to the tunability of their bandgaps between the bandgap of AlN (6.2 eV) and that of GaN (3.4 eV). 

\begin{center}
\begin{figure}
 \includegraphics[width=\linewidth]{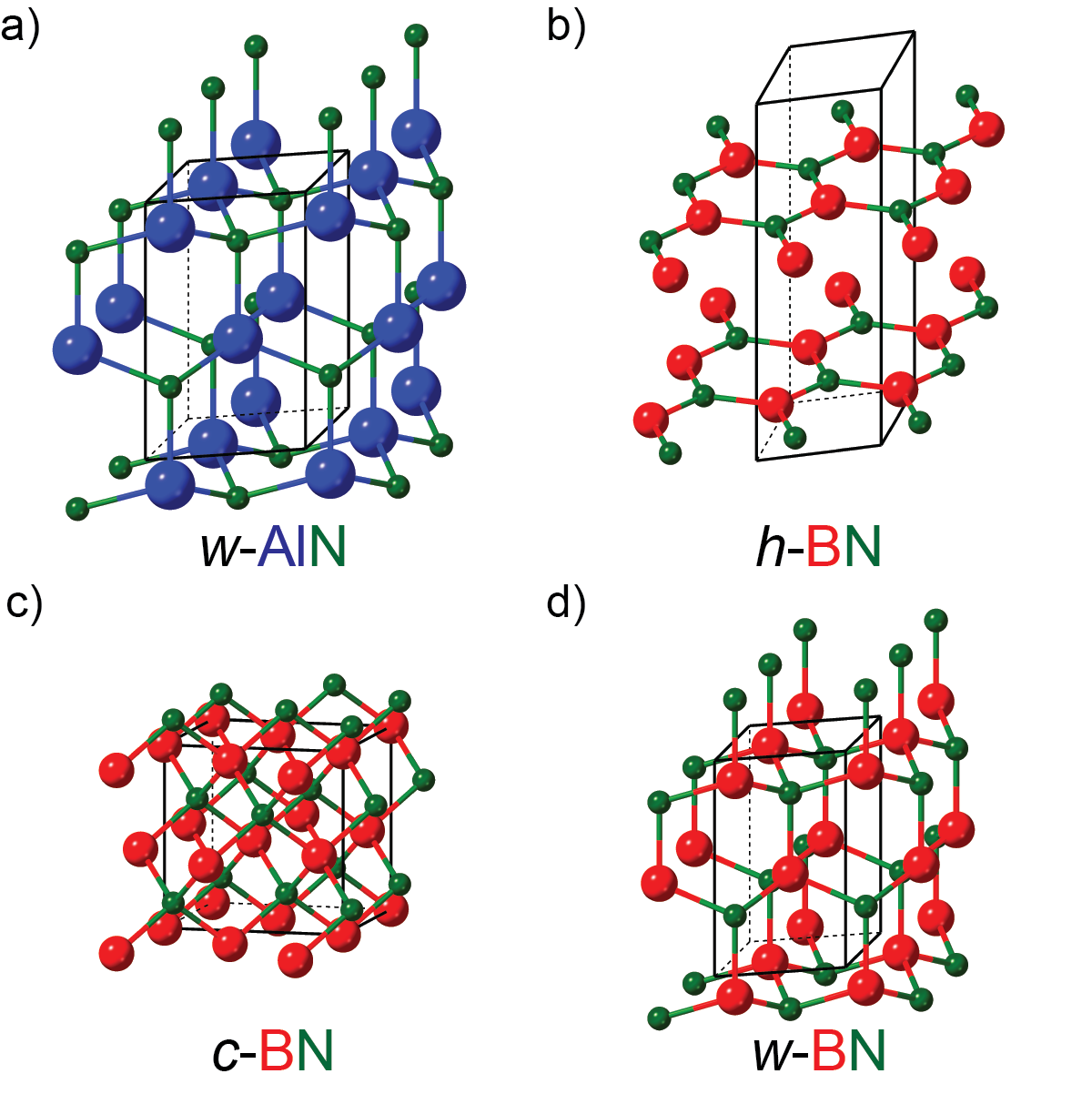}
 \caption{The crystal structures of (a) wurtzite phase of aluminum nitride as well as (b) hexagonal, (c) cubic and (d) wurtzite phase of boron nitride.}
 \label{fig:polymorphs}
\end{figure}
\end{center}

Realization of \baln alloys could allow further tunability of bandgaps beyond what is available via AlGaN alloys. Figure \ref{fig:polymorphs} shows the crystal structures of the wurtzite AlN and BN, as well as the ground state phase of BN, the hexagonal phase. The predicted bandgap of \emph{w}-BN ranges from 5.44 to 7.70 eV\cite{Kudrawiec2020}, much larger than that of GaN (3.4 eV). Additionally, \baln alloys are expected to display high dielectric constants\cite{Hayden2021}. Therefore, \baln may have tunable dielectric constants and high bandgaps without losing the other excellent mechanical and thermal properties. 
However, there are several challenges in the realization of \baln alloys. For example, the lattice mismatch between AlN and BN is very large, 18\%. Moreover, BN has a preference for existing in the hexagonal phase rather than the metastable wurtzite phase, which makes introduction of boron into the wurtzite AlN lattice challenging. Consequently, it leads to the formation of polycrystalline BN phases in \baln alloys and a high density of grain boundaries that can be detrimental to device design and performance\cite{Sun2018}. 
 
\begin{table*}[]
\begin{center}
\begin{tabular}{ c c c c c } 
 \hline
 %E_f values from materialsproject database
 Material & $E_{\text{f}}$ [eV atom$^{-1}$] & $a$ [\AA] & $c$ [\AA] \\ \hline
 \emph{w}-AlN   & -1.584    & 3.128, 3.112\cite{Kudrawiec2020} & 5.016, 4.982\cite{Kudrawiec2020}     \\
 \emph{w}-GaN   & -0.657     & 3.247, 3.189\cite{Kudrawiec2020} & 5.284, 5.185\cite{Kudrawiec2020}    \\
% w-InN   & -0.079     & 3.621, 3.545\cite{Kudrawiec2020} & 5.841, 5.703\cite{Kudrawiec2020}     \\
 \emph{w}-BN   & -1.367     & 2.555, 2.54\cite{Kudrawiec2020}, 2.549\cite{Nagakubo2013} & 4.226, 4.20\cite{Kudrawiec2020}, 4.223\cite{Nagakubo2013}    \\
 \emph{c}-BN   & -1.384     & 3.626, 3.617\cite{Nagakubo2013} & -   \\
 \emph{h}-BN   & -1.461     & 2.500, 2.505\cite{Ooi2006} & 6.426, 6.660\cite{Ooi2006}    \\
 \hline
\end{tabular}
\caption{Comparison of the DFT computed formation energies, $E_{\text{f}}$ and lattice parameters of wurtzite group-III-nitrides and polymorphs of BN. Formation energies are taken from the Materials Project database\cite{Ong2013}.}
\label{table:materials_comparison}
\end{center}
\end{table*}

Despite these challenges, thin-films of \emph{w}-\baln have been successfully grown on AlN and sapphire substrates in very recent years with thicknesses of up to 300 nm and B-fractions up to $x=0.30$\cite{Li2015, Li2017, Sun2018, Tran2020, Vuong2020, Hayden2021, Sarker2020, ZhangQ2022}. In addition, experimental efforts have been made towards understanding their crystal growth and structural properties\cite{Li2017, Sun2018, Sarker2020}. 

Theory and simulations have been used to study the structure of a broader range of B-fractions of \baln alloys. Mainly two classes of methods have been used. The first is the method of cation substitution in the \emph{w}-AlN host lattice\cite{Zhang2017, ZhangQ2022, Shen2017}. The B-fractions that can be studied using this method are dependent on the size of the supercell used to generate the possible \baln structures. Also, as the \baln alloys are highly mismatched alloys, the restriction of the lattice to the wurtzite phase may preclude other structures that could be formed in \baln. Another class of methods is the one used by Ahmed \emph{et al} that uses density-functional theory (DFT) based evolutionary structure searches. This method overcomes the limitations associated with supercell size and constraints set on the symmetry of the crystal to predict the structures. However, due to the high computational resource and time requirement of this method, only three B-fractions, $x=0.25, 0.5, 0.75$, were studied by Ahmed \emph{et al}\cite{Maizhadj2019}.

In this study, we investigate the structure and bonding environment of \baln over the entire boron fraction range using the \emph{ab-initio} cluster expansion (CE) method. In this method, special quasi-random alloys structures are predicted based on a cluster expansion Hamiltonian that is fit to DFT calculated energies. We find ground state phases of \baln at 17 B-compositions with formation energy with respect to AlN and BN between 0.11 to 0.25 eV/atom. We found five metastable phases at $x= 0.333, 0.4, 0.5, 0.6,$ and $0.667$ that are likely to be stabilized by high-temperature growth. We show that the \baln structures deviate from the wurtzite symmetry due to the large lattice mismatch between AlN and BN. Except one ground structure, at $x=0.417$, all ground structures have the cations bonded in the $sp^3$ bonding environment with angles of the tetrahedra $94\degree$ to $141\degree$, and bond lengths between $1.46$ and $1.99$ \AA. The $x=0.417$ structure displays $sp^2$ B bonds that are characteristic of the hexagonal phase of BN. Dynamical stability of the structures was established via phonon spectra simulations. The phonon spectra of all the ground state structures show that they have no imaginary phonon modes and thus are dynamically stable. 

\section{Computational methods}

We employed the cluster expansion methodology as implemented in the Alloy Theoretic Automated Toolkit (ATAT)\cite{avdw:atat,avdw:atat2,avdw:maps,avdw:emc2,avdw:mcsqs} to predict the ground state structures of \baln alloys and compute their total energies. The energy of a system within the CE method can be written as

\begin{center}
\begin{equation}
\begin{split}
  E(\sigma) = J_0 + \sum_i J_i \hat {S_i} (\sigma) + \sum_{j < i} J_{ij} \hat {S_i} (\sigma) \hat{S_j} (\sigma) \\ 
  + \sum_{k < j < i} J_{ijk} \hat {S_i} (\sigma) \hat{S_j} (\sigma) \hat{S_k} (\sigma) + ... 
\end{split}
\label{eq:ce}
\end{equation}
\end{center}

 where $J_i$, $J_{ij}$, and $J_{ijk}$ refer to the CE coefficients for the clusters consisting of one, two, and three atoms, respectively. The value of $\hat{S_i} (\sigma)$ changes depending on if the sites are occupied by Al, B, or N atoms\cite{Wrobel2015}. Together with the constant $J_0$, these CE coefficients were determined from DFT energies of 224 different \baln compounds calculated with the Vienna Ab initio Simulation Package (VASP) package\cite{Kresse1, Kresse2, Kresse3, Kresse4}. Specifically, 97 clusters up to quadruplets were used, the total number of atoms per unit cell were unrestricted, and the wurtzite lattice was used as the lattice system to generate the alloy phases. We obtained a cross-validation (CV) score, which is a measure of the difference between the CE and DFT energies, of $0.012$ eV/atom. The CV score is defined as

\begin{equation} \label{eq:cv}
  (CV)^{2} = n^{-1} \sum_{i=1}^{n} (E_{i} - \hat{E}_{(i)})^{2},
\end{equation}
where $E_{i}$ is the DFT calculated energy (per atom) of structure $i$, $\hat{E}_{(i)}$ is the CE predicted energy of the structure, and $n$ is the number of structures included in the fit. 

All the DFT calculations reported in this study were performed using the Projector Augmented Wave (PAW)\cite{Perdew20} method and the Perdew--Burke--Ernzerhof (PBE)\cite{Blochl1994, Perdew19, Perdew20} exchange-correlation functional as implemented in the VASP package. We have used a plane wave basis set with an energy cut-off of 670 eV and a $k$-grid with $2000$ $k$-points per reciprocal atom for our calculations. The alloy structures are initialized from the AlN lattice and then transformed to supercells using the CE formalism. We then obtained the final structures by relaxing the atomic positions as well as the lattice parameters such that the forces between the ions are less than 0.01 eV/\AA.

The phonon spectra of the ground state structures were calculated using VASP and the phonopy package\cite{phonopy}. For all of the phonon calculations, we have employed a $3 \times 3 \times 3$ supercell for each structure. An energy cut-off of 500 eV was used and the atomic positions were relaxed until the total free energy change between steps was less than $10^{-8}$ eV. For the structures with fewer number of atoms per unit cell ($x=0,0.167,0.25,0.333, 0.5, 0.667, 0.75, 0.833, 1$), density-functional perturbation theory was \\employed\cite{Kresse1, Kresse2, Kresse3,Kresse4, Giannozzi1991,Gonze1997} to obtain the phonon energies. The finite differences method was employed for the alloys with more than 16 atoms per unit cell due to memory constraints. 

\section{Results and Discussion}

\subsection{Formation energies of \baln alloys}

To predict the structure of \baln alloys we calculated the formation energies, $E_{f}$, of the \baln alloys. The $E_{f}$ is a descriptor of thermodynamic stability of a specific structure and is defined as,

\begin{equation} 
  E_{f} = E_{{\mathrm{B}{_x}}\mathrm{Al}{_{1-x}}\mathrm{N}} - x E_\mathrm{BN} - (1-x)E_\mathrm{AlN},
  \label{eq:formationenergy}
\end{equation}
where the total energy per atom of the \baln alloy structures, \emph{w}-BN, and the \emph{w}-AlN are $E_{{\mathrm{B}{_x}}\mathrm{Al}{_{1-x}}\mathrm{N}}$, $E_\mathrm{BN}$, and $E_\mathrm{AlN}$ respectively. 

\begin{center}
\begin{figure}[t!]
 \includegraphics[width=\linewidth]{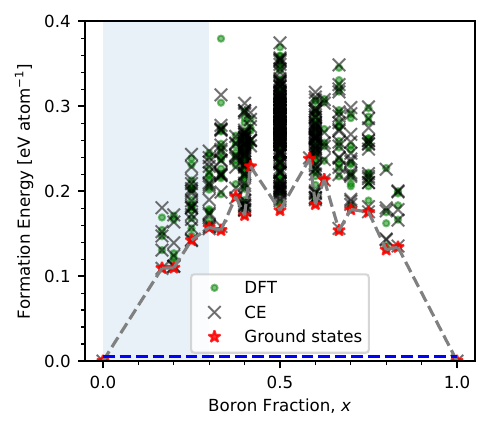}
 \caption{Formation energies, $E_{f}$, of the \baln alloys at different B-fractions, $x$. Green dots represent DFT computed formation energies and gray crosses show CE computed formation energies. The ground state structures are shown as red star symbols. The dashed blue line indicates typical AlGaN formation energies\cite{Bellotti2019}. The shaded region ($x<0.3$) indicates the range of boron fraction in $w$-\baln that has been experimentally synthesized.}
 \label{fig:formationEnergies}
\end{figure}
\end{center}

Figure \ref{fig:formationEnergies} shows the formation energies of the \baln alloys relative to the wurtzite phases of AlN and BN. A similar figure of $E_{f}$ as a function of B-fraction ($x$) with the formation energies of the \baln alloys computed relative to the hexagonal phase of BN instead of wurtzite is included in supplementary information (Figure S1). The CE coefficients described in Eq. \ref{eq:ce} were fitted by considering 8748 structures of which 392 were identified as the best fit and most suitable for DFT structural relaxations. The green circles in Figure \ref{fig:formationEnergies} denote the DFT computed formation energies of the aforementioned 392 structures with the convex hull marked by the gray dashed line. The CE predicted formation energies of these 392 structures are shown with gray crosses in Figure \ref{fig:formationEnergies}. The formation energies of all ground state structures, i.e. structures with the minimum formation energy at each composition $x$, are shown with red stars. Figure \ref{fig:formationEnergies} shows that from our CE and DFT calculations, a total of 17 ground state structures with formation energies in the range 0.11--0.25 eV/atom were predicted.

\begin{center}
\begin{figure}[t!]
 \includegraphics[width=\linewidth]{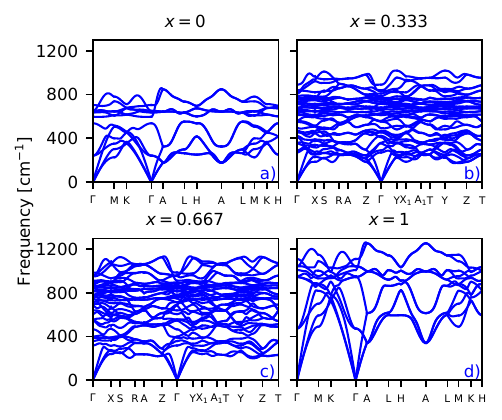}
 \caption{DFT computed phonon spectra of (a) \emph{w}-AlN, (b) B$_{0.333}$Al$_{0.667}$N, (c) B$_{0.667}$Al$_{0.333}$N, and (d) \emph{w}-BN.}
 \label{fig:phonon_grid}
\end{figure}
\end{center}

The formation energies of \baln are more than one order of magnitude larger than that of AlGaN alloys. AlGaN alloy formation energies are typically of the order of $\sim$ 10 meV per atom\cite{Bellotti2019}. The origin of such large formation energies in \baln alloys can be attributed to the lattice mismatch between the \emph{w}-AlN and \emph{w}-BN. Table \ref{table:materials_comparison}, shows the lattice parameters of \emph{w}-AlN, \emph{w}-GaN, and the three phases of BN: \emph{h}-BN, \emph{c}-BN, and \emph{w}-BN. The lattice mismatch between AlN and GaN lattices is quite small, only $\sim$ 4\%, whereas the mismatch is substantial in case of \emph{w}-AlN and \emph{w}-BN, $\sim$ 18\%. As a result, the \baln alloys exhibit high formation energies and thus are challenging to synthesize, unlike AlGaN which is already widely grown for use in power electronics\cite{ColtrinAlGaN2017}. However, it is noteworthy that \baln structures with boron fraction $x \le 0.3$ have been already synthesized\cite{Sun2018, Tran2020, Li2017, Vuong2020}. In this range of $x$ the formation energies of \baln are $\sim$ 100 meV/atom. Formation energies in the range $0.1 - 0.2$ eV/atom correspond to a growth temperatures in the range $1100\degree - 2300\degree$ K, which may be attainable with high temperature growth techniques. 

To further establish the driving factor of the high formation energy, a number of calculations were done with respect to the $x=0.167$ alloy. We investigated the role of lattice strain, the effect of local-bond strains, and the impact of cation-anion tetrahedral distortions. Figure S8 in the SI presents the DFT computed energy estimates for each of these factors. The results show that the formation energies of the \baln alloys are dominated by the effect of the local-bond strains and tetrahedral distortions; whereas the role of the lattice strains are small.

However, it is important to note that formation energy is only one descriptor of the thermodynamic stability of alloys. The growth of semiconducting alloys needs careful consideration of phenomena such as competing phases, substrate effects, method of growth, and defects which are all known to play a large role in the stability and phase purity of the synthesized materials\cite{Li2015, Sun2018, Tolborg2022}. 

\begin{center}
\begin{figure*}[t!]
 \includegraphics[width=\linewidth]{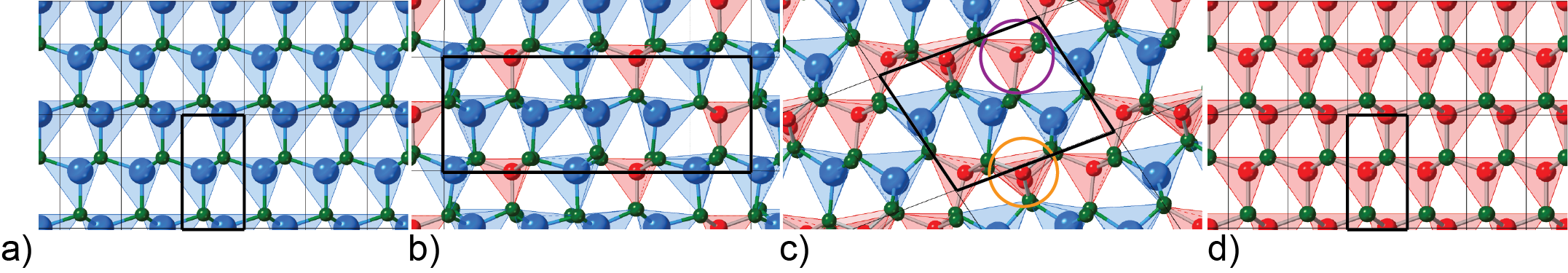}
 \caption{Crystal structures of selected \baln alloys have been shown. The black boxes indicate conventional cells for each alloy. (a) \emph{w}-AlN, (b) B$_{0.30}$Al$_{0.70}$N, (c) B$_{0.417}$Al$_{0.583}$N with an $sp^{2}$ bonded B site circled in orange and an $sp^{3}$ bonded B site circled in purple, and (d) \emph{w}-BN.}
 \label{fig:tetrahedra}
\end{figure*}
\end{center}

To determine the dynamical stability of the ground state structures, we computed the phonon band structure for each alloy structure using density-functional theory. No imaginary modes were found in any of the phonon spectra, indicating that the alloy structures are dynamically stable. 

The calculated $\Gamma$ point phonon frequencies for \emph{w}-AlN and \emph{w}-BN are in excellent agreement with experimental and theoretical values reported in the literature (see SI Table S1). We observe that \emph{w}-BN has much higher phonon frequencies than AlN, indicating a more rigid structure. Therefore, it is expected that higher boron fraction \baln structures will have higher phonon frequencies. Figure \ref{fig:phonon_grid} shows the phonon spectra of \emph{w}-AlN, \emph{w}-BN and some \baln alloys. The phonon spectra of all the ground state structures can be found in the SI Figure S3 and Figure S4. Figure S2 in the SI shows the temperature dependent free energies of formation, which are computed considering the role of vibrational entropy using phonopy\cite{phonopy, Kaczkowski2021}.

Our study shows that there are five metastable \baln phases: at $x= 0.333, 0.4, 0.5, 0.6,$ and $0.667$. These metastable phases can also be observed when the formation energies are computed with respect to \emph{h}-BN instead of the wurtzite phase in Equation \ref{eq:formationenergy} (Figure S1 in the supplementary information). The crystal structures of metastable phases at $x=0.333$ and $x=0.667$ shows well-mixed cation distribution (shown in Figure S5 and Figure S6 in the supplementary information). The metastable phases are likely to remain stable at room-temperature after high-temperature growth and quench to room temperature processes since they lie in a local energy basin. 

In addition to the existence of metastable phases, which appear as local minima at certain boron fraction, we also find a few local maxima in the formation energies e.g., at $x=0.417$ and $x=0.583$ concentrations. Figure \ref{fig:tetrahedra}(c) shows the crystal structures of one such relatively high formation energy \baln alloy. As one can see from Figure \ref{fig:tetrahedra}(c), these structures have alternating cation monolayers and poor cation mixing (see also Figure S6 in the SI). Intermediate boron fraction \baln structures appear to preferentially form alternating cation layers rather than well-mixed structures due to the high lattice mismatch between AlN and \emph{w}-BN.

In the following section \ref{sec3.2}, we further investigate the origin of the non-monotonic behaviour of the formation energy with respect to B-fractions by studying the crystal structure and chemical bonding of these \baln alloys.

\begin{center}
\begin{figure}
 \includegraphics[width=\linewidth]{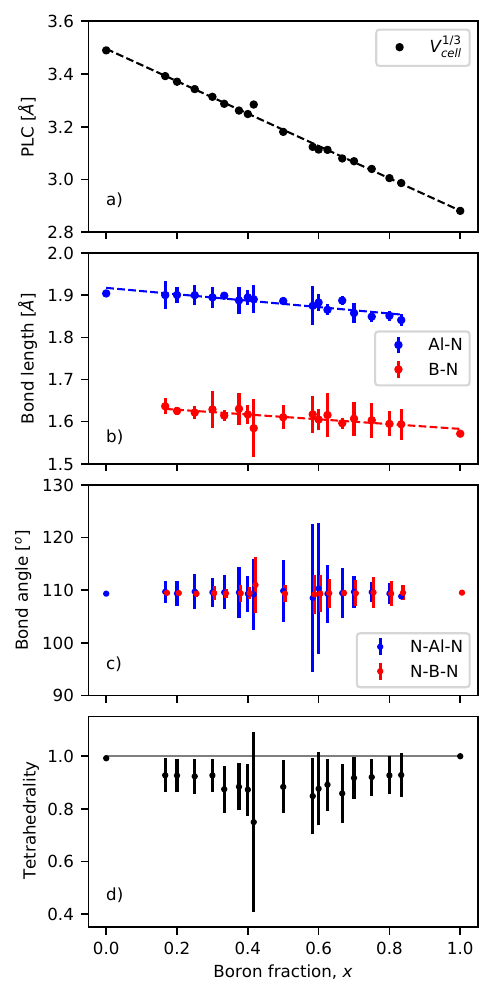}
 \caption{(a) Pseudocubic lattice constant (PLC) of the \baln alloys at various $x$. (b) Average bond lengths calculated for each alloy structure in \AA. (c) Bond angles calculated for each alloy structure are shown in degrees. (d) Average tetrahedrality score calculated over every site in the conventional cell lattice of the \baln alloys, and the solid horizontal line shows a perfect tetrahedrality score of 1. Error bars indicate standard deviations in the values depicted, and dashed lines indicate linear fits to the values.}
 \label{fig:lattice_parameters}
\end{figure}
\end{center}

\subsection{Understanding E$_f$ vs $x$ curve}
\label{sec3.2}

\subsubsection{Pseudocubic lattice constant}

To understand the effect of B-fractions on the formation energy of \baln alloy crystal structures, we first look at the lattice parameters of these alloys. Due to the manner in which the CE, with DFT structural relaxation, computed the ground state structures, there is no restriction that the space group be limited to that of the wurtzite phase. As a result, the space group of the ground state structures at each B-fraction are often different, which makes direct comparison of attributes such as lattice parameters difficult. Table S2 in the SI shows the formation energies, lattice parameters, and space groups for each of the \baln alloy structures. 

Consequently, we use the pseudocubic lattice constant (PLC) to compare the ground state structures' lattice constants. The PLC is defined as the cube root of the volume, $V_{cell}^{1/3}$, of the cell per 4 atoms (similar to the cell of AlN and BN). In other words, the PLC is defined as the normalized cube root of the cell volume and effectively measures the lattice size for structures of different shapes and sizes. The PLC of each structure is shown in Figure \ref{fig:lattice_parameters}(a). The \baln alloy structures closely follow the linear Vegard's law relationship without bowing. Furthermore, the minor deviation at $x=0.417$ is discussed and explained next. 

\subsubsection{Bond lengths and bond angles} 

To further explore the structural properties of \baln at various B-fractions and to explore the anomalous PLC at $x=0.417$, we examined the bond lengths and bond angles in their crystal structures. Figure \ref{fig:lattice_parameters}(b) shows the average Al-N and B-N bond lengths in the conventional cells, with the error bars indicating the standard deviation of their bond lengths. Similar to the PLC, the Al-N and B-N bond lengths of the \baln alloys follow a monotonically decreasing linear relationship with $x$. Minimal deviation is seen in the Al-N bonds from this linear trend. In comparison, the B-N bonds exhibit higher degrees of deviation from the linear trend with larger standard deviations in comparison to the Al-N bonds. The B-N bonds for the $x=0.417$ composition are a clear outlier. The structure of the $x=0.417$ alloy is shown in Figure \ref{fig:tetrahedra}(c) where we can see that the B-N bonds are severely distorted when compared to the purely wurtzite bonds of the other structures.

Figure \ref{fig:lattice_parameters}(c) shows the average N-Al-N and N-B-N bond angles in each ground state alloy structure. Overall, the average bond angles stay constant, although the errors bars, indicating standard deviations, become high for the N-Al-N bonds for intermediate B-fractions. Upon detailed investigation of the crystal structures at these intermediate $x$ values, we find that the high standard deviation in bond angles is caused by the distortion of the tetrahedral bonding environment. All of the ground state crystal structures are depicted in the SI Figures S5 to S7. There is no distinctive feature in the crystal structures that would explain the large standard deviation in the Al-N bonds of $x=0.583$ and $x=0.6$. The N-B-N bond angles and their standard deviation show a similar trend. 

Thus, Figures \ref{fig:lattice_parameters}(a)-(c) show that tetrahedral distortion is largely a result of deviation in the bond angles rather than deviation in the bond lengths. The tetrahedra remain intact but are distorted from regular tetrahedra mostly by changes in the tetrahedral bond angles. At the intermediate values of $x$, one can see rotation and distortion of tetrahedral bonds in the crystal structures in Figures \ref{fig:tetrahedra}(b) and (c). In the following section we will discuss these distortions and their origin in terms of chemical bonding. 

\subsubsection{Tetrahedrality and $sp^{2}$ bonding}

To quantify the distortion in the tetrahedral bonding environment in the \baln alloys, we computed the tetrahedrality of the \baln structures using the \texttt{CrystalNN} method\cite{Pan2021,Zimmermann2020} that is implemented in the open source code \texttt{pymatgen}\cite{Ong2013}. Figure \ref{fig:lattice_parameters}(d) shows the tetrahedrality as a function of $x$ for all the ground state structures. Intermediate amounts of boron decrease the tetrahedrality due to the heightened distortion in the tetrahedra. It appears that the $x=0.417$ structure has the lowest tetrahedrality because of the presence of some trigonal-planar $sp^2$ B-N bonds. 

The presence of $sp^2$ bonding can be explained by noting the fact that the stable phase of BN is the hexagonal phase, which consists of $sp^2$ bonded B atoms in a trigonal planar geometry. Figure \ref{fig:tetrahedra} shows the crystal structure of \baln for $x=0.417$ and $x=0, 0.3,$ and $1$. While both $x=0.417$ and $x=0.30$ structures show large tetrahedral distortions, only the $x= 0.417$ has $sp^2$ bonded B-N. This is shown within the orange circle in Figure \ref{fig:tetrahedra}(c).

Thus overall, the \baln alloys predicted in this work maintain the tetrahedral bonding environment, with the exception of the $x=0.417$ phase. Their space groups and cation arrangements differ but generally their ability to maintain tetrahedral bonding of the wurtzite AlN lattice throughout the entire range of $x$ is observed. 

\section{Discussion and Conclusion}

In summary, we predicted the structures and investigated the bonding environments of \baln over the entire $x=0$ to $1$ boron fraction range using the \emph{ab-initio} cluster expansion method. Ground state phases of \baln were found at 17 B-compositions having formation energies with respect to the wurtzite phases of AlN and BN in the range of 0.11-0.25 eV/atom. The phonon spectra of these \baln alloys indicate that they are dynamically stable across the entire boron fraction range. 

The \baln structures deviate from the ideal wurtzite symmetry due to the large lattice mismatch between AlN and BN. However, the pseudocubic lattice constants of the structures follow the linear Vegard's law. All ground state structures have the cations bonded in the $sp^3$ bonding environment with N-Al-N angles of the tetrahedra between $94$ to $141\degree$ and N-B-N angles of the tetrahedra between $99$ and $125\degree$. Al-N bond lengths are between $1.81$ and $1.99$ \AA\ and B-N bond lengths are between $1.46$ and $1.71$ \AA. One of the ground state structures, at $x=0.417$, also has $sp^2$ B-N bonds that are characteristic to the hexagonal phase of BN.

The growth of \emph{w}-\baln has historically been limited to $<3-4\%$ boron fraction films\cite{ZhangQ2022}, and recent growth has reached up to $x=0.30$\cite{Tran2020}, most likely due to the high formation energies and the observed high structural disorder. Nonetheless, high enough temperatures (1000--2000 K) could facilitate the growth of single-phase \emph{w}-\baln at a higher boron fraction than has previously been reached experimentally. Through this study five metastable phases, at $x=0.333, 0.4, 0.5, 0.6,$ and $0.667$, were identified. They lie in local energy-minima basins and are likely to be stabilized by high-temperature growth. 

\baln alloys are expected to naturally extend the nitride semiconductor material paradigm, and some interesting applications include hetero-barriers for BAlN/AlN majority-carrier structures and quantum barriers for AlN quantum wells (BAlN/AlN/BAlN), which could enable optoelectronics in the ultra-deep UV-range\cite{Tsao2018}. These alloy materials are promising, but implementation into devices requires much more research to understand their electronic and optical properties, such as bandgaps, dielectric behavior, reflection and absorption spectra, breakdown fields, transport properties, and possible doping. This work advances our knowledge of the structures, energies, and bonding in \baln, serving as a foundation for such future studies; and thus eventual integration of \baln in next-generation devices.

\medskip

\textbf{Supporting Information} \par 
Supporting Information is available from the Wiley Online Library or from the author.

\textbf{Acknowledgements} \par %delete if not applicable))
This research was supported by ULTRA, an Energy Frontier Research Center funded by the U.S. Department of Energy (DOE), Office of Science, Basic Energy Science, under Award \# DESC0021230. The authors acknowledge the San Diego Supercomputer Center under the NSF-XSEDE Award No. DMR150006 and the Research Computing at Arizona State University for providing HPC resources. This research also utilized resources from the National Energy Research Scientific Computing Center, a DOE Office of Science User Facility supported by the Office of Science of the U.S. DOE under Contract No. DE-AC02-05CH11231. The authors have no competing interests to declare. 

\bibliographystyle{unsrt}
\bibliography{main}

\begin{thebibliography}{10}

\bibitem{uwbg-semiconductors}
Jeff Tsao, Sakibuddin Chowdhury, M.~Hollis, D.~Jena, N.~Johnson, K.~Jones,
  R.~Kaplar, S.~Rajan, Chris Van~de Walle, E.~Bellotti, C.~Chua, R.~Collazo,
  Michael Coltrin, J.~Cooper, K.~Evans, S.~Graham, T.~Grotjohn, E.~Heller,
  Masataka Higashiwaki, and J.~Simmons.
\newblock Ultrawide-bandgap semiconductors: Research opportunities and
  challenges.
\newblock {\em Advanced Electronic Materials}, 4, 12 2017.

\bibitem{Li2015}
Xin Li, Suresh Sundaram, Youssef~El Gmili, Tarik Moudakir, Fr'ed'eric Genty,
  Sophie Bouchoule, Gilles Patriarche, Russell~D Dupuis, Paul~L Voss, Jean-Paul
  Salvestrini, and A~Ougazzaden.
\newblock {BAlN} thin layers for deep {UV} applications.
\newblock {\em Physica Status Solidi (a)}, 212:745--750, 2015.

\bibitem{Nwigboji2015}
Ifeanyi~H. Nwigboji, John~I. Ejembi, Yuriy Malozovsky, Bethuel Khamala,
  Lashounda Franklin, Guanglin Zhao, Chinedu~E. Ekuma, and Diola Bagayoko.
\newblock Ab-initio computations of electronic and transport properties of
  wurtzite aluminum nitride (\emph{w}-{AlN}).
\newblock {\em Materials Chemistry and Physics}, 157:80--86, 5 2015.

\bibitem{Yu2021}
Ruixian Yu, Guangxia Liu, Guodong Wang, Chengmin Chen, Mingsheng Xu, Hong Zhou,
  Tailin Wang, Jiaoxian Yu, Gang Zhao, and Lei Zhang.
\newblock Ultrawide-bandgap semiconductor {AlN} crystals: growth and
  applications.
\newblock {\em Journal of Materials Chemistry C}, 9:1852--1873, 2 2021.

\bibitem{Kudrawiec2020}
Robert Kudrawiec and Detlef Hommel.
\newblock Bandgap engineering in {III}-nitrides with boron and group {V}
  elements: {T}oward applications in ultraviolet emitters.
\newblock {\em Applied Physics Reviews}, 7(4):041314, 2020.

\bibitem{Hayden2021}
John Hayden, Mohammad~Delower Hossain, Yihuang Xiong, Kevin Ferri, Wanlin Zhu,
  Mario~Vincenzo Imperatore, Noel Giebink, Susan Trolier-Mckinstry, Ismaila
  Dabo, and Jon~Paul Maria.
\newblock Ferroelectricity in boron-substituted aluminum nitride thin films.
\newblock {\em Physical Review Materials}, 5, 4 2021.

\bibitem{Sun2018}
Haiding Sun, Feng Wu, Young~Jae Park, T.~M.~Al Tahtamouni, Che~Hao Liao, Wenzhe
  Guo, Nasir Alfaraj, Kuang~Hui Li, Dalaver~H. Anjum, Theeradetch Detchprohm,
  Russell~D. Dupuis, and Xiaohang Li.
\newblock Revealing microstructure and dislocation behavior in {BAlN}/{AlGaN}
  heterostructures.
\newblock {\em Applied Physics Express}, 11, 1 2018.

\bibitem{Nagakubo2013}
A.~Nagakubo, H.~Ogi, H.~Sumiya, K.~Kusakabe, and M.~Hirao.
\newblock Elastic constants of cubic and wurtzite boron nitrides.
\newblock {\em Applied Physics Letters}, 102, 6 2013.

\bibitem{Ooi2006}
N.~Ooi, V.~Rajan, J.~Gottlieb, Y.~Catherine, and J.~B. Adams.
\newblock Structural properties of hexagonal boron nitride.
\newblock {\em Modelling and Simulation in Materials Science and Engineering},
  14:515--535, 4 2006.

\bibitem{Ong2013}
Shyue~Ping Ong, William~Davidson Richards, Anubhav Jain, Geoffroy Hautier,
  Michael Kocher, Shreyas Cholia, Dan Gunter, Vincent~L. Chevrier, Kristin~A.
  Persson, and Gerbrand Ceder.
\newblock Python materials genomics (pymatgen): A robust, open-source python
  library for materials analysis.
\newblock {\em Computational Materials Science}, 68:314--319, 2013.

\bibitem{Li2017}
Xiaohang Li, Shuo Wang, Hanxiao Liu, Fernando~A. Ponce, Theeradetch Detchprohm,
  and Russell~D. Dupuis.
\newblock 100-nm thick single-phase wurtzite {BAlN} films with boron contents
  over 10\%.
\newblock {\em Physica Status Solidi (B) Basic Research}, 254, 8 2017.

\bibitem{Tran2020}
Tinh~Binh Tran, Che~Hao Liao, Feras Alqatari, and Xiaohang Li.
\newblock Demonstration of single-phase wurtzite {BAlN} with over 20\% boron
  content by metalorganic chemical vapor deposition.
\newblock {\em Applied Physics Letters}, 117, 8 2020.

\bibitem{Vuong2020}
P.~Vuong, A.~Mballo, S.~Sundaram, G.~Patriarche, Y.~Halfaya, S.~Karrakchou,
  A.~Srivastava, K.~Krishnan, N.~Y. Sama, T.~Ayari, S.~Gautier, P.~L. Voss,
  J.~P. Salvestrini, and A.~Ougazzaden.
\newblock Single crystalline boron rich {BAlN} alloys grown by {MOVPE}.
\newblock {\em Applied Physics Letters}, 116, 1 2020.

\bibitem{Sarker2020}
Jith Sarker, Tinh~Binh Tran, Feras Alqatari, Che~Hao Liao, Xiaohang Li, and
  Baishakhi Mazumder.
\newblock Nanoscale compositional analysis of wurtzite {BAlN} thin film using
  atom probe tomography.
\newblock {\em Applied Physics Letters}, 117, 12 2020.

\bibitem{ZhangQ2022}
Qifan Zhang, Qiang Li, Weihan Zhang, Haoran Zhang, Feng Zheng, Mingyin Zhang,
  Peng Hu, Mingdi Wang, Zhenhuan Tian, Yufeng Li, Yuhuai Liu, and Feng Yun.
\newblock Phase transition and bandgap engineering in {B$_{1-x}$Al$_x$N}
  alloys: {DFT} calculations and experiments.
\newblock {\em Applied Surface Science}, 575:151641, 2 2022.

\bibitem{Zhang2017}
Muwei Zhang and Xiaohang Li.
\newblock Structural and electronic properties of wurtzite {B$_x$A$_{1-x}$N}
  from first-principles calculations.
\newblock {\em Physica Status Solidi (B) Basic Research}, 254, 8 2017.

\bibitem{Shen2017}
Jimmy~Xuan Shen, Darshana Wickramaratne, and Chris G. Van~De Walle.
\newblock Band bowing and the direct-to-indirect crossover in random {BAlN}
  alloys.
\newblock {\em Physical Review Materials}, 1, 11 2017.

\bibitem{Maizhadj2019}
H.~{Maiz Hadj Ahmed}, H.~Benaissa, A.~Zaoui, and M.~Ferhat.
\newblock Exploring new insights in baln from evolutionary algorithms ab initio
  computations.
\newblock {\em Physics Letters A}, 383(13):1385--1388, 2019.

\bibitem{avdw:atat}
A.~van~de Walle, M.~D. Asta, and G.~Ceder.
\newblock {T}he {A}lloy {T}heoretic {A}utomated {T}oolkit: {A} user guide.
\newblock {\em Calphad}, 26:539--553, 2002.

\bibitem{avdw:atat2}
A.~van~de Walle.
\newblock {M}ulticomponent multisublattice alloys, nonconfigurational entropy
  and other additions to the {A}lloy {T}heoretic {A}utomated {T}oolkit.
\newblock {\em Calphad}, 33:266--278, 2009.

\bibitem{avdw:maps}
A.~van~de Walle and G.~Ceder.
\newblock Automating first-principles phase diagram calculations.
\newblock {\em J. Phase Equilib.}, 23:348--359, 2002.

\bibitem{avdw:emc2}
A.~van~de Walle and M.~D. Asta.
\newblock Self-driven lattice-model monte carlo simulations of alloy
  thermodynamic properties and phase diagrams.
\newblock {\em Model. Simul. Mater. Sc.}, 10:521, 2002.

\bibitem{avdw:mcsqs}
A.~van~de Walle, P.~Tiwary, M.~M. de~Jong, D.~L. Olmsted, M.~D. Asta, A.~Dick,
  D.~Shin, Y.~Wang, L.-Q. Chen, and Z.-K. Liu.
\newblock Efficient stochastic generation of special quasirandom structures.
\newblock {\em Calphad}, 42:13--18, 2013.

\bibitem{Wrobel2015}
Jan~S. Wróbel, Duc Nguyen-Manh, Mikhail~Yu Lavrentiev, Marek Muzyk, and
  Sergei~L. Dudarev.
\newblock Phase stability of ternary fcc and bcc {Fe-Cr-Ni} alloys.
\newblock {\em Physical Review B - Condensed Matter and Materials Physics}, 91,
  1 2015.

\bibitem{Kresse1}
G.~Kresse and J.~Hafner.
\newblock Ab initio molecular dynamics for liquid metals.
\newblock {\em Physical Review B}, 47:558, 1993.

\bibitem{Kresse2}
G.~Kresse and J.~Hafner.
\newblock Ab initio molecular-dynamics simulation of the
  liquid-metal-amorphous-semiconductor transition in germanium.
\newblock {\em Physical Review B}, 49(20):14251, 1994.

\bibitem{Kresse3}
G.~Kresse and J.~Furthm\"{u}ller.
\newblock Efficiency of ab-initio total energy calculations for metals and
  semiconductors using a plane-wave set.
\newblock {\em Computational Materials Science}, 6:15, 1996.

\bibitem{Kresse4}
G.~Kresse and J.~Furthm{\"{u}}ller.
\newblock Efficient iterative schemes for ab initio total-energy calculations
  using a plane-wave basis set.
\newblock {\em Physical Review B}, 54:11169--11186, 1996.

\bibitem{Perdew20}
J.~P. Perdew, K.~Burke, and M.~Ernzerhof.
\newblock Erratum generalized gradient approximation made simple.
\newblock {\em Phys. Rev. Lett.}, 78:1396, 1997.

\bibitem{Blochl1994}
P.~E. Blochl.
\newblock Projector augmented-wave method.
\newblock {\em Physical Review B}, 50:17953, 1994.

\bibitem{Perdew19}
J.~P. Perdew, K.~Burke, and M.~Ernzerhof.
\newblock Generalized gradient approximation made simple.
\newblock {\em Physical Review Letters}, 77:3865, 1996.

\bibitem{phonopy}
A~Togo and I~Tanaka.
\newblock First principles phonon calculations in materials science.
\newblock {\em Scripta Materialia}, 108:1--5, Nov 2015.

\bibitem{Giannozzi1991}
Paolo Giannozzi, Stefano~De Gironcoli, Pasquale Pavone, and Stefano Baroni.
\newblock Ab initio calculation of phonon dispersions in semiconductors.
\newblock {\em Physical Review B}, 43, 1991.

\bibitem{Gonze1997}
Xavier Gonze and Changyol Lee.
\newblock Dynamical matrices, born effective charges, dielectric permittivity
  tensors, and interatomic force constants from density-functional perturbation
  theory.
\newblock {\em Physical Review B}, 55:10355--10368, 4 1997.

\bibitem{Bellotti2019}
Alexandros Kyrtsos, Masahiko Matsubara, and Enrico Bellotti.
\newblock First-principles study of the impact of the atomic configuration on
  the electronic properties of {Al$_x$Ga$_{1-x}$N} alloys.
\newblock {\em Physical Review B}, 99, 1 2019.

\bibitem{ColtrinAlGaN2017}
Michael~E. Coltrin and Robert~J. Kaplar.
\newblock Transport and breakdown analysis for improved figure-of-merit for
  {AlGaN} power devices.
\newblock {\em J. Appl. Phys.}, 121(5):055706, Feb 2017.

\bibitem{Tolborg2022}
Kasper Tolborg, Johan Klarbring, Alex~M. Ganose, and Aron Walsh.
\newblock Free energy predictions for crystal stability and synthesisability.
\newblock {\em Digital Discovery}, 1:586--595, 2022.

\bibitem{Kaczkowski2021}
Jakub Kaczkowski and Iwona Płowaś-Korus.
\newblock The vibrational and thermodynamic properties of {CsPbI$_3$}
  polymorphs: An improved description based on the {SCAN} meta-{GGA}
  functional.
\newblock {\em Journal of Physical Chemistry Letters}, 12:6613--6621, 7 2021.

\bibitem{Pan2021}
Hillary Pan, Alex~M. Ganose, Matthew Horton, Muratahan Aykol, Kristin~A.
  Persson, Nils~E.R. Zimmermann, and Anubhav Jain.
\newblock Benchmarking coordination number prediction algorithms on inorganic
  crystal structures.
\newblock {\em Inorganic Chemistry}, 60:1590--1603, 2 2021.

\bibitem{Zimmermann2020}
Nils~E.R. Zimmermann and Anubhav Jain.
\newblock Local structure order parameters and site fingerprints for
  quantification of coordination environment and crystal structure similarity.
\newblock {\em RSC Advances}, 10:6063--6081, 2020.

\bibitem{Tsao2018}
J.~Y. Tsao, S.~Chowdhury, M.~A. Hollis, D.~Jena, N.~M. Johnson, K.~A. Jones,
  R.~J. Kaplar, S.~Rajan, C.~G.~Van de~Walle, E.~Bellotti, C.~L. Chua,
  R.~Collazo, M.~E. Coltrin, J.~A. Cooper, K.~R. Evans, S.~Graham, T.~A.
  Grotjohn, E.~R. Heller, M.~Higashiwaki, M.~S. Islam, P.~W. Juodawlkis, M.~A.
  Khan, A.~D. Koehler, J.~H. Leach, U.~K. Mishra, R.~J. Nemanich, R.~C.N.
  Pilawa-Podgurski, J.~B. Shealy, Z.~Sitar, M.~J. Tadjer, A.~F. Witulski,
  M.~Wraback, and J.~A. Simmons.
\newblock Ultrawide-bandgap semiconductors: Research opportunities and
  challenges.
\newblock {\em Advanced Electronic Materials}, 4, 1 2018.

\end{thebibliography}

\end{document}